\documentstyle{article}

\begin{document}

\author{Michael Fowler}
\title{Quantum Integrable Systems in One Dimension }
\date{September 1993 }
\maketitle

\section{Introduction}

The field of integrable systems is a vast one that has grown almost
explosively in recent years. It is not remotely possible to review all the
developments in the field in one hour, so I am restricting my attention to
some particular topics. I shall only consider one dimensional quantum
systems, or 1+1 as it is usually said, including the time dimension, but
necessarily therefore including some discussion of the statistical mechanics
of two dimensional lattice systems (vertex models) which share a common
mathematical structure with the 1+1 quantum systems, and have been the key
to much of the understanding of the quantum systems.\cite{baxter} \cite
{sutherland86} \cite{korepin93}

This is still far too big a field to review in any detail in one hour, so I
shall focus on some generalizations of the spin one-half chain with
nearest-neighbor coupling, for which Bethe first set down his celebrated
ansatz\cite{bethe31}. The simplest generalization, to fully anisotropic
coupling, the so-called $XYZ$ model, has an amazingly rich mathematical
structure, parts of which are still being elucidated. A second
generalization, which has been very fashionable lately, is to extend the
coupling beyond nearest neighbors, specifically to a long range
inverse-square type coupling between spins one-half. This also turns out to
be an integrable system, although the wave functions are not of the Bethe
ansatz type.\cite{haldane88} \cite{shastry88}

How relevant are these models to real physical systems? The simplest
isotropic spin one half $XXX$ Heisenberg antiferromagnet is a good
Hamiltonian for the quasi-one-dimensional system CPC, and one of the early
experimental vindications of the Bethe ansatz was the confirmation by
neutron scattering\cite{cpc} that it correctly predicted the observed
elementary excitation spectrum, in contrast to the standard spin wave theory
used at the time. An appropriate continuum limit of the $XYZ$ chain gives
the sine-Gordon model, and the $XXZ$ chains are good representations of
known systems. The closely related one-dimensional Hubbard model has been
widely used to describe one-dimensional conductors, and may be relevant to
some high temperature superconductors. One of the great successes of the
Bethe ansatz has been the solution of the Kondo problem\cite{andrei83}. The
Luttinger liquid model\cite{haldane80}, first used to analyze low energy
properties of the $XXZ$ spin chain, is proving valuable in studying
one-dimensional mesoscopic conducting systems\cite{kanefisher}. In fact,
Bethe ansatz models are very useful for one-dimensional conducting ring
systems having a threading magnetic field, because the Bohm-Aharonov phase
can be incorporated as a twisted boundary condition which, it turns out,
does not spoil the integrability. Another very fruitful interaction in
recent years has been that between Bethe ansatz methods and conformal field
theory, discussed below. Finally, the inverse-square systems\cite
{sutherland72} are cousins of the Bethe ansatz systems. They do not have
Bethe ansatz wavefunctions, but their energy levels are given by Bethe
ansatz like equations of a rather simple kind. These systems are closely
related to the important edge states in the quantum Hall effect, and also,
surprisingly, to level distributions in quantum chaotic systems\cite
{simonsleealt}.

\section{Quick review of Bethe ansatz}

The simplest Bethe ansatz system to visualize is the ``billiard balls on a
line'' problem---classically, imagine $N$ perfectly elastic billiard balls
on a ring, say, to give periodic boundary conditions. The initial set of
momenta $k_1,...,k_N$ is conserved in subsequent collisions. The quantum
mechanical version, solved long ago by Lieb and Liniger\cite{liebliniger},
has $N$ bosons interacting through repulsive delta function potentials. The
two particle problem is equivalent to a single particle interacting with a
static delta-function potential, and the standard boundary condition at the
potential leads to a two-particle wavefunction which is a sum of two plane
wave terms with the momenta permuted,
\begin{equation}
\label{one}\psi (x_1,x_2)=e^{i(k_1x_1+k_2x_2)}+e^{i\theta
_{12}}e^{i(k_2x_1+k_1x_2)}
\end{equation}
for $x_1<x_2$, and boson symmetry determines the wavefunction for $x_1>x_2$.

The phase shift term is found to be:
\begin{equation}
\label{two}e^{i\theta _{12}}=-\frac{c-i(k_1-k_2)}{c+i(k_1-k_2)}
\end{equation}
where $c$ is the strength of the repulsive delta-function potential. This is
a simple Bethe ansatz wavefunction. Notice that it {\it vanishes} if $%
k_1=k_2 $.

Surprisingly, this wavefunction generalizes in a simple way to $N$
particles,
\begin{equation}
\label{three}\psi (x_1,x_2,...,x_N)=\sum a(P)e^{i\sum k_{P_j}x_j}
\end{equation}
for $x_1<x_2<...<x_N$, with other orderings given by the boson symmetry.
Here $a(P)$ is a product of phase shift factors for each pair of transitions
needed to get the ordering $P$ from the ordering $1,...,N$. It follows
immediately from the form of the phase shift term above that {\it all }the $%
k_i$ must be distinct, or the wavefunction vanishes identically. The total
energy and total momentum are easily seen to be given by $E=\sum k_i^2$, $%
K=\sum k_i$, from applying the $E,K$ operators in regions where no two $x_i$%
's coincide.

The actual values of the $k_i$ appearing in the wavefunction are determined
by the boundary conditions, just as they are for free particles. For
periodic boundary conditions, the total change in the phase of the wave
function on taking a particle around the system must be $2\pi I$, where $I$
is an integer. This phase is part kinetic, from the $e^{ikx}$ term, and part
phase shifting from passing the other particles so the total phase shift
\begin{equation}
\label{four}k_jL+\sum_{i\neq j}\theta _{ji}=2\pi I_j
\end{equation}
Now writing the phase shift in terms of the momenta, we get a set of coupled
nonlinear equations for the $k_i$'s for a given set of quantum numbers $I_i$%
. Since the $k_i$ must all be distinct, the ground state is fermi-sea like,
with a set of $I_j$ equal to $0,$ $\pm 1,\pm 2,...$ . Elementary excitations
are given by gaps (holes) in the sequence of integers, or single integers
above the sea (particles). It is found that for an {\it attractive} delta
function potential, the ground state for $N$ particles is a {\it string} of
pure imaginary momenta having uniform spacing $ic$. Actually, this
attractive boson gas is a pathological system in the thermodynamic limit
where all the particles bind together infinitely tightly. However, these
strings of bound momenta prove to be a common (usually nonpathological)
feature of Bethe ansatz systems, as discussed below.

Let us now turn to the spin one-half $XXZ$ chain having Hamiltonian:
\begin{equation}
\label{five}H_{XXZ}=\sum_{j=1}^N\left( S_j^xS_{j+1}^x+S_j^yS_{j+1}^y+\Delta
S_j^zS_{j+1}^z\right)
\end{equation}
where we shall restrict our attention to the gapless regime, $|\Delta |<1$,
and write $\Delta =\cos \gamma $. The eigenstates of the Hamiltonian are
described by Bethe ansatz wavefunctions having the same general structure as
in the boson model, the bosons being replaced by magnons. A magnon is
created by applying the operator $a_k^{+}=\sum e^{ikn}S_n^{+}$ to the
all-spins-down ``ferromagnetic'' state. It is convenient to parameterize the
magnon momenta $k_j$ in terms of so-called rapidity variables $v_j$:
\begin{equation}
\label{six}e^{ik_j}=\frac{\sinh (v_j+i\gamma /2)}{\sinh (v_j-i\gamma /2)}
\end{equation}
The reason for the change of variables is that the magnon-magnon phase shift
is a function of rapidity difference. It was {\it not }a function of
momentum difference---the lattice spoils Galilean invariance. The equations
are much easier to handle in the new variables. The boundary condition for
allowed rapidities analogous to (4) (or, rather, its exponential) is:
\begin{equation}
\label{seven}\left( \frac{\sinh (v_j+i\gamma /2)}{\sinh (v_j-i\gamma /2)}%
\right) ^N=-\prod_{i=1}^n\frac{\sinh (v_j-v_i+i\gamma )}{\sinh
(v_j-v_i-i\gamma )}
\end{equation}

One significant difference from the (repulsive) boson system is that the
magnon rapidities given by solving these equations include in general some
grouped in strings of certain lengths, the allowed lengths depending on $%
\gamma $, and also showing a very complicated system size dependence for
finite systems.

One of the most important applications of the Bethe ansatz method is finding
thermodynamic properties. Roughly speaking, this is done by summing over an
ensemble of systems having the same macroscopic quantum number density with
appropriate entropic weighting factors. For the spin chains, the presence of
so many string-like excitations makes this a formidable task even in the
infinite limit\cite{takahashi72}, and for finite systems, where string
lengths vary with system size, it is even more tricky. Fortunately, new
techniques have recently been discovered which circumvent these difficulties%
\cite{saw} \cite{ddv92}. We shall discuss them below.

\section{Luttinger liquids and conformal fields}

One real problem with the above analysis is that, although all the energy
levels of the Bethe ansatz system can be found and the wavefunctions written
down, it is very difficult in practice to find matrix elements of operators
between these states, and hence connect the known excitation spectrum with
what is observed in, say, neutron scattering from a spin chain. The first
significant step in this direction was made by Luther and Peschel\cite
{lutherpeschel}, who argued that the low energy excitations of the gapless $%
XXZ$ spin chain could be described using the Luttinger model, solved by
Mattis and Lieb\cite{liebmattis}. These ideas were extended and put on a
firm footing by Haldane\cite{haldane80}, who coined the term Luttinger
liquid. Essentially, the method is to transform the magnons to spinless
fermions using the Jordan-Wigner transformation, whereupon the $XX$ term is
seen to be a kinetic fermion hopping contribution to the Hamiltonian, and
the $\Delta $ term is a four-fermion interaction. The fermion
energy-momentum curve is then linearized near the ``fermi surface'' so that
density fluctuations can be represented by bosons. With this approximation,
the whole low-energy Hamiltonian is quadratic in the boson operators, and
can be diagonalized and solved. The fermion creation operator is found to be
closely related to an exponential function of boson creation operators. The
critical exponents can be found in terms of bulk susceptibilities which can
be calculated by conventional Bethe ansatz methods. This Luttinger liquid
theory was a forerunner of, and is in fact a special case of, modern
conformal techniques.

As is discussed further in the next section, the 1+1 dimensional Bethe
ansatz systems are intimately related to two dimensional statistical
systems. Essentially, they are analytic continuations of each other from
time to imaginary time. Thus techniques developed for one can be readily
applied to the other.

Macroscopic thermodynamic systems at the critical point have been long known
to be scale invariant. Polyakov\cite{polyakov} first pointed out that they
are also {\it conformally} invariant, that is, invariant under any
transformation that is locally a scale change plus a rotation. In two
dimensions, this constraint dramatically reduces possible varieties of
critical behavior. The conformal transformations of the plane, under which
the system must be invariant, correspond to the set of all analytic
functions, and the generators of these transformations are the elements of
the so-called Virasoro algebra. Representations of this algebra,
corresponding to possible critical systems, are labelled by a central charge
$c$ and other parameters usually denoted $\Delta _i$. For the statistical
systems, the central charge measures the finite-size dependence of the free
energy (or the low temperature specific heat for the spin chain), and the
other parameters measure correlation exponents for both systems. It was
shown by Friedan, Qiu and Shenker\cite{fqs} that unitarity considerations
only allow certain values of $c$ and the $\Delta _{i}$, and these values
determines possible excitation energies of the system, in particular the
low-lying states. But these are precisely the energy levels that can be
evaluated fairly straightforwardly using the Bethe ansatz, so by this
backdoor method we can use Bethe ansatz results to learn about correlation
functions without trying to evaluate matrix elements! A great deal of
numerical work has been done to classify the conformal theories
corresponding to continuum limits of integrable spin chain systems.\cite
{frahmyf}

The Luttinger liquid technique (with some conformal ideas added) has been
applied successfully to the Hubbard model by Frahm and Korepin\cite{frahmk90}%
. This is a more complicated situation, in that there are{\it \ two }sets of
low energy excitations, charge and spin density waves, having different
speeds. This means that conformal methods need to be extended. The basic
tool used is the dressed charge matrix, a generalized susceptibility, the
elements of which can be calculated using the Bethe ansatz. This immediately
gives asymptotic correlation functions in terms of coupling strength and
magnetic field. Recently, Kane and Fisher\cite{kanefisher} have applied
Luttinger liquid techniques to predicting conductivity in a one-dimensional
system with barriers.

\section{Transfer matrices and analyticity}

Our growing understanding of the $XXZ$ (and fully anisotropic $XYZ$) spin
one half chains has come about in large part through their connections with
the six (and eight) vertex models. These are classical two dimensional
statistical models defined on a square lattice, such that on each bond there
is an arrow which can point either way. The energy, and therefore the
statistical weight, of each vertex depends on its arrow configuration. In
the six vertex model, originally conceived to describe hydrogen-bonded
crystals (the arrow direction indicates the end of the bond where the
hydrogen atom sits) electrostatic energy considerations dictate that each
vertex has two and only two hydrogen atoms nearby, so two arrows point in
and two out at each vertex. This gives six possible vertices, three
arrow-reversed pairs. (The eight vertex model allows all in or all out
arrows.) Taking arrow reversed pairs to have the same energy, there are
three distinct statistical weights, usually denoted $a$, $b$, and $c$. The
free energy of these lattice systems can be found by transfer matrix
techniques---the matrix is defined between adjacent rows of $N$ vertical
bonds, each matrix element giving the combined statistical weight of all
allowed vertex configurations on the row of horizontal bonds sandwiched
between the two adjacent rows of specified vertical bonds. This is a $%
2^N\times 2^N$ matrix with elements having row and column labels
corresponding to the $2^N$ possible spin configurations on each of two
successive rows of vertical bonds. Denoting this matrix by $T$, the free
energy of the $N\times N$ system is given by $TrT^N$, so the matrix
multiplication sums over all possible vertical bond configurations, with the
appropriate Boltzmann weighting factor supplied by the matrix. For large $N$%
, the trace is dominated by the largest eigenvalue of $T$, so this gives the
free energy. Finding correlation functions of the system requires knowledge
of the next leading eigenvalues.

Now the transfer matrix $T$ operates on a space of $N$ up or down arrows,
which is the {\it same} as the configuration space of the spin-one-half
chain Hamiltonian. This was first appreciated by Lieb\cite{liebice}, who
solved the ice problem by using a Bethe ansatz wavefunction for the leading
eigenstate of the transfer matrix. However, the relationship between $T$'s
and $H_{XXZ}$'s is not one-to-one. Taking vertex weights with all arrows
reversed to be equal, and factoring out an overall normalization, we still
have a{\it \ two}-parameter family of vertex models, with transfer matrices
conveniently labelled $T(v,\gamma )$, where $T(v,\gamma )$ has the same set
of eigenstates (but not of course eigenvalues) as the member $H_{XXZ}(\gamma
)$ of the one-parameter $XXZ$ spin-chain family. That is to say, if we take $%
\gamma $ as fixed, there is a one-parameter family of transfer matrices $T(v)
$ which all commute with $H_{XXZ}(\gamma )$, and in fact they all commute
with each other, since they have the same set of eigenstates. In the
notation of Kl\"umper et al.,\cite{klumpbatpearce} the ratio of the three
vertex statistical weights $a:b:c$ becomes $\sin (\gamma /2+iv):\sin (\gamma
/2-iv):\sin \gamma $, and it can easily be shown that for the particular
parameter values $v=\pm i\gamma /2$, the transfer matrices reduce to simple
shift operators (because $a$ or $b$ become zero), and their logarithmic
derivatives at these points are just the $XXZ$ Hamiltonian,
\begin{equation}
\label{eight}T(\mp (i\gamma /2-v))=e^{\pm iP-vH_{XXZ}+O(v^2)}
\end{equation}
{}.

It might appear at this point that introducing this family of commuting
transfer matrices does not obviously look like progress in understanding the
properties of the $XXZ$ spin chain. In fact, though, the existence of the
family $T(v)$ makes it evident that the $XXZ$ chain is integrable, in the
sense of having a sequence of higher conserved quantities---i.e., beyond
total momentum and energy---one could, for example, find the higher
logarithmic derivatives of $T(v)$ at $v=\pm i\gamma /2$. However, the real
utility of the $T(v)$ arises from their analytic properties. As Baxter makes
clear throughout his classic book, the analyticity of $T(v)$ as a function
of $v$ constrains the properties of these systems, and renders them
solvable, in a truly amazing way. As an example of the connection between
the Bethe Ansatz equations and analyticity, we mention Baxter's result that
each eigenvalue $\Lambda (v)$ of the transfer matrix satisfies the equation:
\begin{equation}
\label{nine}\Lambda (v)q(v)=\Phi (v-i\gamma /2)q(v+i\gamma )+\Phi (v+i\gamma
/2)q(v-i\gamma )
\end{equation}
with $\Phi (v)=(\sinh v)^N,$and $q(v)=\prod_{j=1}^n\sinh (v-v_j)$. The
rapidities $v_j$ satisfy the Bethe ansatz equations $p(v_j)=-1$ where
\begin{equation}
\label{ten}p(v)=\frac{\Phi (v-i\gamma /2)q(v+i\gamma )}{\Phi (v+i\gamma
/2)q(v-i\gamma )}
\end{equation}

These are of course the same equations as in (7) above, but in the context
of the properties of the transfer matrix eigenvalue as a function of $v$, we
see that the zeros of $q(v)$ coincide with zeros of the right hand side of
equation (9), ensuring that $\Lambda (v)$ has no poles. Furthermore, for the
largest eigenvalue, it is known that the $v_i$are all real, from which it
follows that $\Lambda (v)$ is analytic and nonzero in a strip of width $%
\gamma $ centered on the real axis. This is a very strong constraint, and
makes it possible to use Cauchy's theorem to derive a nonlinear integral
equation for the function $1/p(x-i\gamma /2)$, which is {\it exact} even for
{\it finite} systems. Although the equation cannot be completely solved, it
can be used to find the $1/N$ corrections to the eigenvalues for large
systems. This means we can evaluate critical exponents analytically\cite
{klumpbatpearce} \cite{baxter} without directly solving the Bethe ansatz
equations.

As mentioned earlier, the standard method for evaluating thermodynamic
properties of Bethe Ansatz systems leads to formidably difficult
calculations involving large numbers of string-like excitation densities,
even the completeness of the set of states is sometimes hard to establish.
Fortunately, it has been realized quite recently that the free energy can
also be evaluated more directly using the analyticity properties. One needs
to evaluate $Tr(e^{-\beta H_{XXZ}})$. Writing this in discretized fashion as
\begin{equation}
Tr\left( e^{-\frac \beta N.H_{XXZ}}\right) ^N
\end{equation}
at first glance it appears to be dominated by the largest eigenvalue of the
operator in the bracket. This is of course incorrect, because as $N$ goes to
infinity, all the eigenvalues degenerate to unity. The trick to getting
around this is to use crossing symmetry\cite{saw} \cite{ddv92}. A neat
presentation was given recently by Kl\"umper\cite{klumper93}. From equation
(8) above,
\begin{equation}
T\left( i\gamma /2-\beta /N\right) .T\left( -i\gamma /2+\beta /N\right)
=e^{-\frac \beta N.H_{XXZ}}
\end{equation}
Raising this to the $N^{th}$ power, one has a product of transfer matrices
for a six-vertex model having two {\it different} vertex strengths,
appearing on alternate rows, a horizontally striped lattice. To evaluate
this in the large $N$ limit, we switch to the equivalent product of {\it %
column} transfer matrices. The column transfer matrices are all identical,
but down each column the vertex strengths alternate, as it crosses all the
stripes. The eigenvalues of this{\it \ inhomogeneous} transfer matrix do
{\it not} all collapse to unity as $N$ becomes large---the limit is well
defined, and leads to an integral equation from which one can find the
leading eigenvalue. Therefore, by this method we can find the thermodynamic
behavior of the spin chain in a very direct way, without the headache of
summing over many different string excitations, and concerns about
completeness. The new approach proves to be very effective computationally.

\section{Inverse-square type systems}

At this point, we back up to our $N$ billiard balls on a line model (meaning
point bosons), but this time add an inverse-square repulsive potential
between pairs of billiard balls. This is a famous classical integrable
system, solved by Calogero\cite{calogero}, and others, using a Lax pair of
matrices. One result that emerged was that if initially the balls are far
apart with momenta $p_1,p_2,...,p_N$ then the final state has them far apart
with the same set of momenta. This suggested to Sutherland\cite{sutherland72}
the idea of an {\it asymptotic Bethe ansatz} for the quantum system. The
wavefunction is certainly not of the standard Bethe ansatz form of a sum of
products of plane waves, because the smoothly varying potential energy
between particles will give functions of smoothly varying wavelength,
destroying the plane wave property. Nevertheless, the dilute system will be
close to plane wave form in most of configuration space, and if the
scattering phases are factorizable so that diffraction does not occur, the
Bethe ansatz might make sense in this dilute limit. Actually, the phase
shifting is particular simple for the $1/r^2$ potential, since there is no
scale, it is independent of relative momentum, except for the sign. To be
precise, for a potential of strength $2\lambda (\lambda -1)/r^2$, the phase
shift is $\pi \lambda k/\left| k\right| .$ This makes the Bethe ansatz
equations like (4) above extremely easy to solve! Of course, by going to
equations like (4), we have switched to periodic boundary conditions, but it
turns out this can be accomplished by replacing the inverse-square potential
by the periodic version, $2\lambda (\lambda -1)/\sin ^2r$, without spoiling
the integrability. In fact, the ground state wavefunction for this potential
is known exactly. It is $\prod_{i<j}\left| \sin (x_i-x_j)\right| ^\lambda $,
quite unlike a Bethe ansatz wavefunction. Yet, remarkably, the asymptotic
Bethe ansatz equations give the exact ground state energy of the system even
in this manifestly nondilute limit. They also give all the excitation
energies correctly---in other words, the full spectrum and the
thermodynamics. Note that for $2\lambda =1,2$ or $4$ the ground state
correlation function is exactly the distribution of eigenstates of random
matrices found by Dyson\cite{dyson62} for orthogonal, unitary and symplectic
ensembles.

The high degeneracy of the spectra indicates the presence of new types of
symmetries\cite{haldane91}. There has been much recent work on the algebraic
structures of these systems, see for example the review of Bernard\cite
{bernard}. An illuminating interpretation has been suggested by Haldane\cite
{haldane92}, who argues that the magnon excitations of the inverse-square
spin one-half chain are actually a gas of free anyons, specifically spin
one-half semions, so the dynamics of the system follows entirely from the
statistical ``interaction''.

As a simple example of the power of the algebraic approach, we present here
a new exchange operator formalism invented by Polychronakos\cite
{polychronakos}, which gives an easy way to prove the integrability of some
of these systems. For $N$ particles on a line with positions and momenta $%
x_i,p_i$ he introduces an operator $M_{ij}$ that permutes the positions and
momenta of particles $i,j$, then defines a generalized momentum operator
\begin{equation}
\pi _i=p_i+i\sum_{i\neq j}V(x_i-x_j)M_{ij}
\end{equation}
with $V$ as yet unspecified. He then requires the Hamiltonian to have
free-particle form in the $\pi _i$'s,
\begin{equation}
H=\frac 12\sum \pi _i^2
\end{equation}
which gives
\begin{equation}
H=\frac 12\sum p_i^2+\frac 12\sum_{i\neq j}\left(
iV_{ij}(p_i-p_j)M_{ij}+V_{ij}^{^{\prime }}M_{ij}+V_{ij}^2\right) -\frac
16\sum V_{ijk}M_{ijk}
\end{equation}
with
\begin{equation}
V_{ijk}=V_{ij}V_{jk}+V_{jk}V_{ki}+V_{ki}V_{ij}
\end{equation}
and $M_{ijk}$generates cyclic permutations, $M_{ijk}=M_{ij}M_{jk}$.

Requiring the Hamiltonian to have only two-body potentials, $V(x)$ must
satisfy
\begin{equation}
V(x)V(y)+V(y)V(z)+V(z)V(x)=W(x)+W(y)+W(z), x+y+z=0
\end{equation}
The commutator of the $\pi _i$'s is
\begin{equation}
\left[ \pi _i,\pi _j\right] =\sum V_{ijk}(M_{ijk}-M_{jik})
\end{equation}
At this point, it is very easy to establish the integrability of the
Calogero system, in other words, to find the $N$ independent constants of
motion of the system. Assume $V(x)=l/x$. Then the Hamiltonian has the form
\begin{equation}
H=\frac 12\sum p_i^2+\sum_{i>j}\frac{l(l-M_{ij})}{(x_i-x_j)^2}
\end{equation}
and since on the boson system $M_{ij}=0$, this is just the Calogero system.
Now for $V(x)=l/x$, from (16) $V_{ijk}=0$, so from (18) the $\pi _i$ all
commute, and, trivially, so do the symmetric quantities
\begin{equation}
I_n=\sum \pi _i^n
\end{equation}
Assuming now $V(x)=l\cot ax$ gives the periodic system. The $\pi _i$ no
longer all commute, but by adding $l\sum_{j\neq i}M_{ij}$ to each we
generate a new set of $I_n$ that do commute, so again we establish
integrability with a minimum of effort!

In fact, this route leads directly to the integrability of the
Haldane-Shastry spin chain system. All we have to do is to give the bosons a
spin of one-half, and take the classical (infinitely heavy boson) limit.
(This is a mathematical model, so we do not worry about spin and
statistics.) On these bosons, the operator $M_{ij}$ above is essentially
equivalent to a spin exchange operator. In the classical limit, the bosons
form a lattice ground state, and the spin Hamiltonian is identical to the
Haldane-Shastry model---the momentum term disappears in the expression for $H
$ above, establishing the integrability of the spin chain\cite{fmp}. The
analogous argument applied to the Calogero system of bosons confined by an
external harmonic potential gives a Haldane-Shastry type system in which the
spins are located at points corresponding to zeros of an Hermite polynomial.
Again, the eigenstates are highly degenerate.\cite{frahmpoly}

Another inverse-square integrable system that has been extensively
investigated recently\cite{gebhardruck} is the Hubbard model with long-range
hopping:
\begin{equation}
H=\sum t_{m,n}c_{m\sigma }^{+}c_{n\sigma }+U\sum n_{m\uparrow
}n_{m\downarrow }
\end{equation}
where
\begin{equation}
t_{m,n}=\frac{it(-1)^{m-n}}{\frac L\pi \sin \frac \pi L}
\end{equation}
This system also has highly degenerate energy levels, given by Bethe ansatz
like equations which are much more tractable than those for the standard
Hubbard model, so this is an interesting easy model for exploring properties
of one-dimensional conductors.

Finally, we should at least mention the intriguing connection between
integrability and random matrices, although this will be discussed in more
detail by Altshuler at this Conference. As stated above, the ground state
wavefunction for the periodic inverse-square boson system with interaction
parameter $\lambda $, $\prod_{i<j}\left| \sin (x_i-x_j)\right| ^\lambda $,
is for $2\lambda =1,2$ or $4$, the distribution function of eigenvalues
found by Dyson for orthogonal, unitary and symplectic random matrix
ensembles. Can any meaning be given to the time-dependent correlation
functions of the boson system in terms of random systems? Simons, Lee and
Altshuler\cite{simonsleealt} explored how the eigenvalues of a random system
(electronic energy levels in a disordered conductor) respond to a steadily
growing external potential (or a twisting boundary condition). For the
disordered conductor, they were able to use supersymmetry methods developed
by Efetov\cite{efetov} to evaluate eigenvalue correlations. They argue that
for a potential
\begin{equation}
H=H_0+\lambda V
\end{equation}
where $H_0$ is a random potential, containing many noninteracting electrons,
writing $\lambda ^2=2it$, the movement of the eigenvalues as $\lambda $ is
increased exactly mirrors the dynamics of the bosons with the inverse-square
type interaction. Perhaps it is a suitable note to end on that one of the
most ordered systems imaginable, an integrable system with many higher
symmetries reflected in its simple spectrum, is intimately connected
mathematically to a totally disordered system.

\end{document}